\documentclass[useAMS,usenatbib]{mn2e}

\usepackage{graphics}
\usepackage{psfig}
\def\acollata{\frac{a_{\rm coll}}{a_{\rm ta}}}

\def\acoll{a_{\rm coll}}
\def\ata{a_{\rm ta}}

\title[Dark energy and the evolution of spherical overdensities]
{Dark energy and the evolution of spherical overdensities}
\author[Cathy Horellou and Joel Berge]{Cathy Horellou\thanks{E-mail:
horellou@oso.chalmers.se} and Joel Berge\thanks{Current address: 
Service d'Astrophysique, CEA Saclay, F-91191 Gif-sur-Yvette, France}\\
Onsala Space Observatory, 
Chalmers University of Technology, SE-439 92 Onsala, Sweden}
\begin{document}

\date{Accepted . Received ; in original form }

\pagerange{\pageref{firstpage}--\pageref{lastpage}} \pubyear{}

\maketitle

\label{firstpage}

\begin{abstract}
We use the non-linear spherical model 
in cold dark matter (CDM) cosmologies with dark energy   
to investigate the effects of dark energy on 
the growth of structure 
and the formation of virialised structures. 
We consider dark energy models with a constant equation of state parameter $w$. 
For $-1<w<-1/3$, clusters form earlier and are more concentrated in 
quintessence than in $\Lambda$CDM models, 
but they form later and are less concentrated 
than in the corresponding open model with the same matter density and  
no dark energy. 
We point out some confusion in the literature around the 
expression of the collapse factor (ratio of the radius of the sphere at virialisation
to that at turn-around) derived from the virial theorem.   
We use the Sheth \& Tormen extension of the Press-Schechter framework to calculate
the evolution of the cluster abundance in different models 
and show the sensitivity of the cluster abundance to both the amplitude of the
mass fluctuations, $\sigma_8$, and the $\sigma_8-w$ normalisation,  
selected to match either the cosmic microwave background observations or the 
abundance of X-ray clusters. 
\end{abstract}


\section{Introduction}

The cosmological picture that has emerged from a host of 
recent observations is that of a spatially flat universe 
dominated by an exotic component with positive energy
density and negative pressure, the dark energy, which is 
responsible for the observed accelerated cosmic expansion. 

The nature of dark energy remains unknown. 
It could be Einstein's cosmological constant 
or the manifestation of a scalar field slowly rolling down
its potential, as in quintessence models  
(\citealt{peeblesratra88}, 
\citealt{ratrapeebles88}, 
\citealt{caldwell}, 
see \citealt{peeblesratra03} 
for a review). 
The equation of state,  
$w=p_X/\rho_X$, where $p_X$ is the pressure and 
$\rho_X$ the density of the dark energy 
(hereafter referred to as the X-component), describes
how the X-component evolves as the universe expands. 
Determining the equation of state
will help reveal the nature of 
dark energy and discriminate among models, 
as the cosmological constant is characterised by 
a constant $w=-1$ whereas in quintessence models $w$ can be different  
from $-1$ and vary with cosmic time. 
$w$ is directly related to the shape of the potential of the
quintessence field. It is a fundamental parameter that remains poorly
constrained observationally, partly because of degeneracies with other
parameters, mainly $\Omega_0$ and the r.m.s. of the amplitude of fluctuations 
within 8~$h^{-1}$Mpc, $\sigma_8$ 
(e.g., \citealt{lokas04}).

Dark energy not only affects the expansion rate of the background and
the distance-redshift relation,  
but also the growth of structure. 
The collapse of overdense regions due to gravitational instability 
is slowed down by the Hubble drag due to the expansion. 
This effective friction depends on the expansion of the background. 
The formation rate of haloes, their evolution and their final
characteristics are modified. 
Dark energy is therefore expected to have an impact on observables such as 
cluster number counts and 
lensing statistics due to intervening concentrations of mass on the
line-of-sight of background sources.  

The abundance of rich clusters of galaxies can be used to constrain the cosmological
model and the properties of dark energy. 
This has been discussed first for pure cosmological constant models 
(e.g., \citealt{lahav}, 
\citealt{lilje92}, 
\citealt{lacey}, 
\citealt{vianaliddle}, 
\citealt{eke})
and then for models with dark energy where $w$ could be different from $-1$. 
In a seminal paper 
\cite{wang} have discussed the evolution of the cluster abundance in general quintessence models 
with negative pressure.  
Following up on that paper, \cite{lokas01}, \cite{basilakos} have examined    
cluster formation in models with constant $w$.  
\cite{mota} have selected particular potentials for the quintessence field. 
\cite{battye} have investigated the constraints on dark energy from future
Sunyaev-Zeldovich surveys. 
\cite{lokas04} have used $N$-body simulations in flat models with constant $w$ to 
measure cluster mass functions. 

In this paper, we first discuss the linear evolution of perturbations 
in dark energy models with constant $w$. Then,  
we use the non-linear spherical model (``top-hat'' model
first developed by \citealt{gunn}). 
Rather than using the approximation given by \cite{wang} which is strictly valid 
for $w=-1$, we calculate explicitely the potential energy associated with the dark 
energy component inside the collapsing sphere and the ratio of the radius of the sphere 
at virialisation to the turn-around radius.
We follow the evolution of overdense regions as they first expand with
the background before collapsing and settling into virialised
structures. 
We then calculate the abundance of 
massive clusters in the \cite{press74} framework.
We use the \cite{st99} extension which has been shown to be in remarkable agreement 
with results from $N$-body  
simulations of hierarchical structure formation. 

We use five cosmological models, 
all with $H_0=70$ kms$^{-1}$Mpc$^{-1}$: 
three dark energy models with the same energy content: 
$\Omega_0$=0.3, $\Omega_{X,0}$=0.7, but 
that differ in their value of $w$: 
$\Lambda$CDM with $w=-1$, 
a $w=-0.8$ model, 
a $w=-0.6$ model, 
and for comparison 
an Einstein-de Sitter model with $\Omega_0=1$ and 
an open model with $\Omega_0=0.3$. 

\section{The background cosmology}

We consider a cosmological model with two components: 
(1) a non-relativistic component which comprises all forms of matter 
(luminous and dark) that clusters under the action of gravity, 
and (2) a uniform X-component with negative pressure 
which does not cluster at the scales of
interest. 
In general, quintessence models have a component that is spatially inhomogeneous
but does not cluster on scales less than 100~Mpc.
The Friedmann equation that describes 
the evolution of the scale factor $a$ with cosmic time 
is modified to include 
the effect of the X-component: 

\begin{equation}
{\left( \frac{H(a)}{H_0} \right)}^2 = 
{\Omega_0 \frac{\rho(a)}{\rho_0}
         +\Omega_k a^{-2}
	 +\Omega_{X,0} \frac{\rho_X(a)}{\rho_{X,0}}
	}
\end{equation}
where $H(a)=\dot a/a$ is 
the Hubble parameter, 
$\Omega$ are the density parameters 
(ratios of the energy density to 
the critical density 
$\rho_{\rm crit}= \frac{3H^2}{8\pi G}$)
for the matter, the curvature 
($\Omega_k =  1-\Omega_0 -\Omega_{X,0}$) 
and the X-component respectively. 
The parameters are 
indexed by 0 when they refer to the present time. 
 
The energy density of the X-component varies as
\begin{equation}
\frac{\rho_X(a)}{\rho_{X,0}} = a^{-n}
\end{equation}
with $n=3(1+w)$.  

Different $n$ and $w$ correspond to different types of energy densities, with 
$n=0$ ($w=-1$) corresponding to the cosmological constant. 
Note that $n$ doesn't need to be either integer or constant. It may 
vary with cosmic time, as in kinessence models. 
In the following we consider
only models with constant $n$ and $w$.  

Note also that for $n=2$ ($w=-1/3$), the dark energy term can be
combined with the curvature term in the Friedmann equation; 
$\Omega_{X,0}$ cancels out and  
the expansion is the same as for a $\Omega_0$ universe. 
Thus a flat universe with $w=-1/3$ 
behaves like an open universe without dark energy and 
with the same $\Omega_0$. 
Cosmic acceleration (i.e., the existence of an inflexion point 
in the $a(t)$ curve) requires $n<2$.  

The Friedmann equation can be written more simply as 
\begin{equation}
\dot a = \frac{H_0}{f(a)}
\label{eq-friedmann}
\end{equation}
where 
\begin{equation}
f(a)= a^{-1}\left[ \Omega_0 a^{-3} 
                  + (1-\Omega_0 - \Omega_{X,0})a^{-2}
		  + \Omega_{X,0} a^{-n} 
             \right]^{-1/2}. 
\label{eq-fa}
\end{equation}

\begin{figure}
\psfig{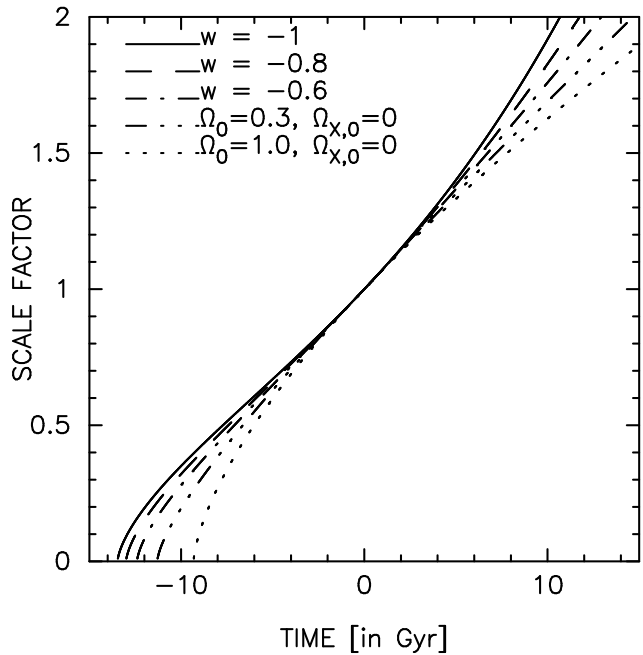}
\caption{Evolution of the scale factor in the different models. 
In quintessence models, the evolution of the scale factor is intermediate
between that in a $\Lambda$CDM model (continuous line) and that in an
open model (dashed-dotted-dotted-dotted line).
}
\label{at}
\end{figure}

Integrating the Friedmann equation, one obtains the variations of 
the scale factor with cosmic time in different cosmologies (see Fig.~\ref{at}). 
Universes with a dark energy component are older than with $\Omega_X=0$ 
and give perturbations more time to grow. 
For the currently favoured parameters
the evolution of the scale factor in quintessence models is
intermediate between 
that of a cosmological constant model
($w=-1$) and of an open universe. 

The density parameters vary with redshift $z=1/a-1$ as 
\begin{equation}
\Omega(z) = \frac{\Omega_0 (1+z)^3}{E^2(z)} \,\, ; 
\qquad 
\Omega_X(z) = \frac{\Omega_{X,0}(1+z)^n}{E^2(z)}
\end{equation}
where $E(z) = \frac{H(z)}{H_0}$. 
In the case of the cosmological constant commonly denoted $\Lambda$, 
\begin{equation}
\Omega_X = \lambda = \frac{\Lambda}{3H^2}\, . 
\end{equation}

\section{Growth of linear perturbations}\label{sect-growth}

\begin{figure}
\psfig{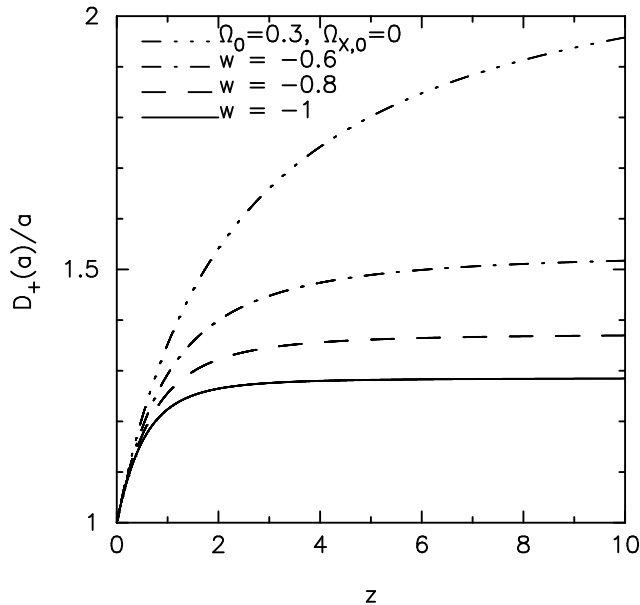}
\caption{Evolution of the growth factor normalised to the
present time and divided by the scale factor $a$ as a 
function of redshift for different cosmological models. 
In an Einstein-de Sitter universe $D_{+}/a=1$ at all redshifts. 
$D_{+}/a$ is highest in the open model. 
In quintessence models it takes intermediate values between 
the open model and the $\Lambda$CDM model. 
}
\label{dz}
\end{figure}

In linear theory, the evolution of the density contrast 
of a perturbation 
$\delta=(\rho_{\rm pert}-\rho)/\rho$ 
is governed by 
\begin{equation}
\ddot\delta + 2\frac{\dot a}{a}\dot\delta -4\pi G\rho\delta = 0  
\end{equation}
which can be solved numerically. 
Analytical solutions  
have been given for $\lambda_0=0$ 
(e.g. \citealt{peebles}), 
$\Omega_0+\lambda_0=1$ 
(\citealt{bildhauer}; 
\citealt{chernin}; 
\citealt{lokaspreprint}). 
A way to estimate the growth factor numerically for any $(\Omega_0,\lambda_0)$
pair has been given by \cite{hamilton}. 
\cite{basilakos} provides a generic approximation formula for $D_{+}(z)$ 
as a function of $\Omega_0, \Omega_X$ and $w$ 
that agrees 
with the
approximations given by \cite{lahav} and \cite{carroll}  
for the $\Lambda$CDM case.  
\cite{linderjenkins} discuss the influence of the equation of state 
(including a time-varying $w$) on the growth of structure. 

It is common practice to normalise the growth factor to its value at 
the present time. 
Fig.~\ref{dz}  shows the variations of the growth factor normalised at 
$z=0$ and divided by
$a$ as a function of redshift 
for the different cosmological models. 
In an Einstein-de Sitter universe $D_{+}/a$ is always equal to 1, 
and structures can form at any redshift. 
The curves show that the growth factor is highest in an open universe, 
and it takes intermediate values for decreasing $w$ up to  
$w=-1$, the cosmological constant case. 
A $\Lambda$CDM universe has a higher growth rate than an 
Einstein-de Sitter universe throughout its history, but
the growth rate drops at redshifts below $\sim1/\Omega_0$.  
For $-1<w<-1/3$ the decrease in the growth rate occurs at higher and
higher redshifts and over a large range of redshifts as $w$ increases. 
In other words, structure growth will be stronger at early times
in an open universe compared to a $\Lambda$CDM universe, 
with intermediate rates in quintessence models. 
But structure growth is reduced earlier in open and quintessence
models, although it is still significantly higher than 
in $\Lambda$CDM models. 

\section{Non-linear evolution of a spherical
overdensity}\label{sect-nonlin}

The equation of motion of a spherical shell in the presence of dark energy is: 
\begin{equation}
\frac{\ddot{r}}{r}= -\frac{4\pi G}{3}\left( \rho_{\rm cluster} + \rho_{\rm X,eff} \right) 
\label{eq-motion}
\end{equation}
where $\rho_{\rm cluster}$ is the time-varying density inside the
forming cluster, and 
\begin{equation}
\rho_{\rm X,eff} = \rho_X + 3 p_X = (n-2) \rho_X
\label{eq-rhoxeff}
\end{equation}
is the effective energy density of
the X-component. 
$\rho_{\rm X,eff}$ is constant for the cosmological constant; 
in a universe with a dark energy component with $n=2$ ($w=-1/3$), 
perturbations evolve in the same way as in a universe with the same
$\Omega_0$ and no dark energy. 
Note that for $n\neq0$ and $n\neq 2$ ($w\neq -1$ and $w\neq -1/3$), 
the density of the X-component inside the
overdensity patch is dependent on the evolution of the background. 
This makes it impossible to integrate directly the equation of motion
and cast it into a first-order energy equation. 
The second-order
equation of motion has to be integrated taking into account the 
time variation of the dark energy.

We shall follow the evolution of a spherical overdense region 
with some initial overdensity. 
At early times, it expands along with the Hubble flow 
and density perturbations grow proportionally to the scale factor. 
If the initial overdensity exceeds a critical value, 
the overdense region will break away from the
general expansion and 
go through three phases:
\begin{enumerate}
\item{expansion up to a maximum radius;}
\item{collapse;}
\item{virialisation.}
\end{enumerate}
We call $z_{\rm ta}$ the redshift at maximum expansion (turn-around)
and $z_{\rm coll}$ the redshift at which the sphere suddenly
virialises, 
$r_{\rm ta}$ and $r_{\rm vir}$ the corresponding radii of the
sphere. 
Ideally, the sphere should collapse down to an infinitely
small radius, but it is assumed that this is prevented by 
the growth of small inhomogeneities. 
Let us discuss in turn the different
phases of evolution of a collapsing sphere.

\subsection{Expansion up to $r_{\rm ta}$}\label{sect-expansion}

Let us calculate the overdensity of the forming cluster at turn-around
\begin{equation}
\zeta = \left( \frac{\rho_{\rm cluster}}{\rho_{\rm bg}}\right)_{z=z_{\rm ta}} \,
.
\end{equation}
Let us define 
\begin{equation}
x = \frac{a}{a_{\rm ta}} \qquad {\rm and} \qquad
y=\frac{r}{r_{\rm ta}}
\end{equation}
where $a_{\rm ta}$ is the scale factor of the background when the
perturbation reaches turn-around. 
For a flat universe, the evolution of the background
and that of the perturbation are governed by the two following 
equations: 
\begin{equation}
\dot{x} = H_{\rm ta} \Omega_{\rm ta}^{1/2} [\Omega(x) x]
^{-1/2} 
\label{eq-diff1}
\end{equation}
 \begin{equation}
\ddot{y} = - \frac{H_{\rm ta}^2\Omega_{\rm ta}}{2} \left[
\frac{\zeta}{y^2} + (n-2)\frac{\Omega_{\rm X,ta}}{\Omega_{\rm ta}} 
\frac{y}{x^n} 
\right]  
\label{eq-diff2}
\end{equation}
where we have used the fact that 
the mass of the forming cluster is 
conserved: 
\begin{equation}
\rho_{\rm cluster}r^3 = 
\rho_{\rm cluster,ta} r_{\rm ta}^3 \, .
\end{equation}
Note that the second term 
in eq.~(\ref{eq-diff2}) vanishes in two cases: for perturbations in a 
$n=2$
universe and for the ones in a universe  
with $\Omega_X=0$.
In the $n=0$ (the cosmological constant) case also, 
the evolution of the perturbation is independent of that
of the background.  

\begin{figure}
\psfig{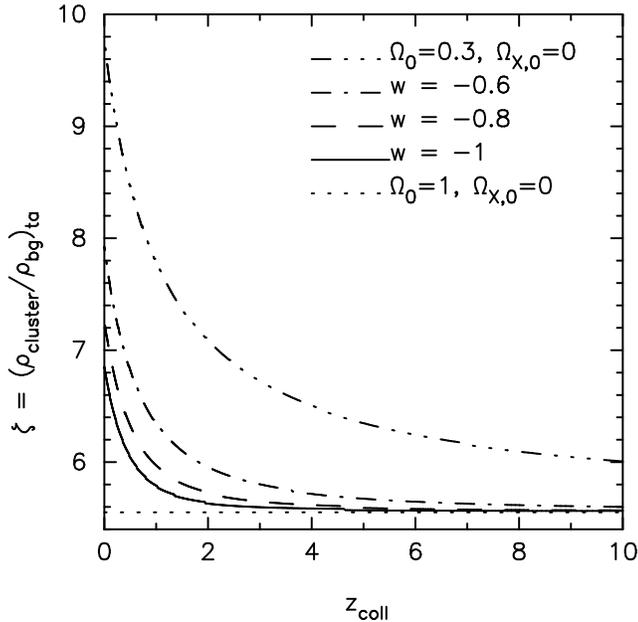}
\caption{Overdensity at turn-around versus collapse redshift 
in different cosmological models. 
Clusters are denser at turn-around in quintessence models than
in the $\Lambda$CDM model. 
Quintessence models are intermediate between 
the $\Lambda$CDM model and the open model. 
$\zeta\simeq5.6$ in an Einstein-de Sitter universe. 
}
\label{fig-zetazcoll}
\end{figure}

$\zeta$ can be determined by integrating the above
differential equations using the boundary conditions 
$(dy/dt)_{x=1}=0$ and $(y)_{x=0}=0$. 
The variations of $\zeta$ with collapse redshift are shown in 
Fig.~\ref{fig-zetazcoll}.  
\cite{wang} provide a fitting formula for $\zeta$ 
as a function of $\Omega_{\rm ta}$ and $w$ for a spatially flat 
Universe. Perturbations collapsing at a certain redshift 
are denser at turn-around relative to the
background in a universe with quintessence than in a $\Lambda$CDM
model. 
At high redshift, $\zeta$ tends toward the fiducial value of 5.6 
for an Einstein-de Sitter universe (as $\Omega$ tends toward 1 and
$\Omega_X$ toward 0). 

\begin{figure}
\psfig{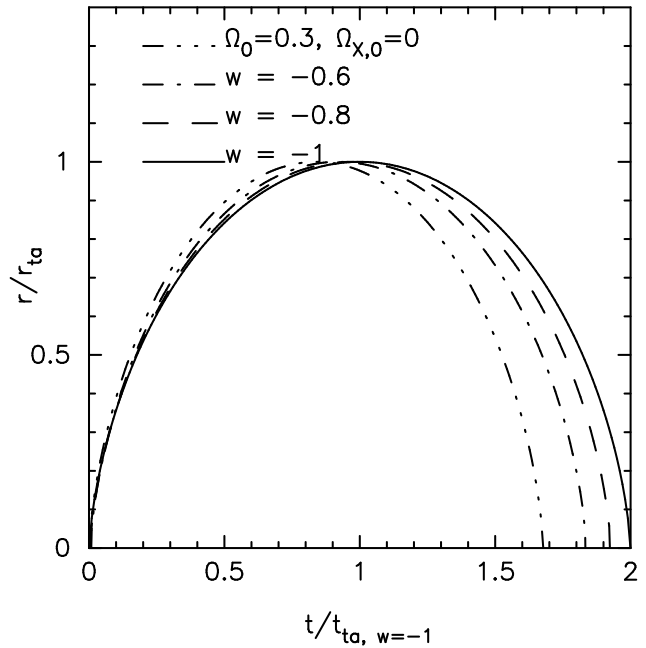}
\caption{Evolution of a perturbation collapsing at the present time 
($z_{\rm coll}=0)$ in different cosmologies. 
The radius $r$ has been normalised to the turn-around radius $r_{\rm ta}$. 
The $x$-axis shows the time normalised to the turn-around time for the
$\Lambda$CDM model. 
Perturbations reach turn-around and collapse earlier in the quintessence model 
than in the $\Lambda$CDM model 
and even earlier in the OCDM model.
}
\label{fig-rrta}
\end{figure}

Fig.~\ref{fig-rrta} shows the evolution of a perturbation collapsing
now
($z_{\rm coll}=0$) in different cosmological models. 
Perturbations reach turn-around and collapse earlier in the quintessence models
than in the $\Lambda$CDM model, and even earlier in the OCDM model.  


\subsection{Collapse and virialisation}\label{sect-collapse}
\begin{figure}
\psfig{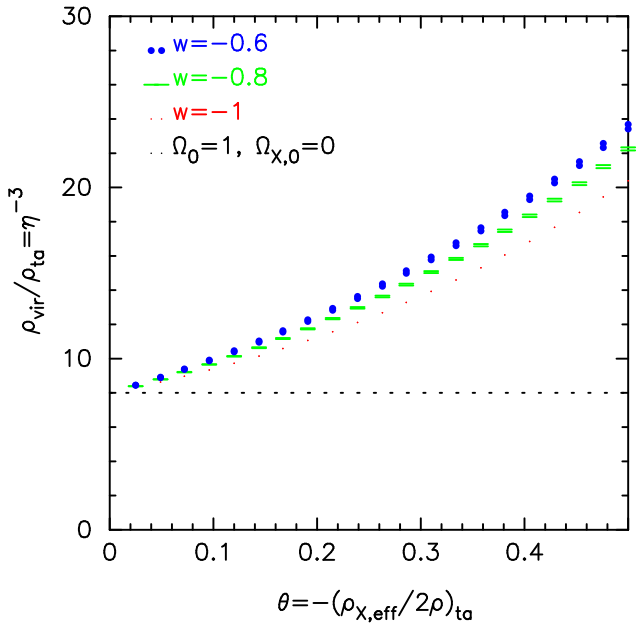}
\caption{Ratio of the density of the virialised sphere 
to the density at turn-around $\rho_{\rm vir}/\rho_{\rm ta} = \eta^{-3}$ 
as a function of $\theta$, the strength of the dark energy at
turn-around,  for different
models. 
The dotted line corresponds to an Einstein-de Sitter model 
(density ratio of 8). 
Values are shown for increasing $\acollata$ from 1.60 to 1.70 by 
steps of 0.02. 
}
\label{rhotheta}
\end{figure}

The virial theorem can be expressed in the general form
$T=pU/2$ for a system with a potential energy of the form
$U\propto R^p$ and kinetic energy $T$ (\citealt{landau}). For the cosmological
constant and the quintessence, the potential energy is
proportional to $R ^2$, and we can therefore use the
virial theorem.
Combining the virial theorem and the conservation of energy
one obtains a relation between the potential energies of the collapsing sphere
at turn-around and at the time it virialises:
\begin{equation}
U_{\rm G,ta} + U_{\rm X,ta}= {{1}\over{2}}U_{\rm G,vir} + 2 U_{\rm X,vir}
\label{eq-virial}
\end{equation}
where $U_G$ is the potential energy of the matter 
($U_G=-{{3}\over{5}}{{GM^2}\over{R}}$)
and $U_X$ is the potential energy of the quintessence inside the collapsing
sphere.

The potential energy associated with the quintessence $U_X$ can be obtained from 
the Poisson equation with the pressure term:
\begin{equation}
\triangle \Phi_X = 4\pi G (\rho_X + 3p_X) 
\end{equation}
which yields:
$\Phi_X = (n-2) \rho_X {{2\pi G}\over{3}} R^2 $
and
\begin{equation}
U_X =  (n-2) \rho_X {{4 \pi G M}\over{10}} R^2.
\end{equation}
Replacing the energies by their expressions in Eq.~(\ref{eq-virial}), 
we can obtain a relation between the 
collapse factor, defined as 
\begin{equation}
\eta=\frac{r_{\rm vir}}{r_{\rm ta}} 
\end{equation}
and the magnitude of the dark energy at turn-around. 
Following \cite{iliev}, let us define a quantity $\theta$ to  
describe 
the importance of the X-component 
relative to the gravitational attraction of the matter at maximal
expansion:
\begin{equation}
\theta=-\Bigl(\frac{\rho_{X,{\rm eff}}}{2\rho}\Bigr)_{\rm ta} 
\end{equation}
where $\rho_{X,{\rm eff}}$ is defined in eq.~(\ref{eq-rhoxeff}). 

Noting that 
\begin{equation}
\left(\frac{\rho_{\rm X,eff}}{\rho}\right)_{\rm vir}=
2\theta\eta^3\left(\acollata\right)^{-n} 
\end{equation}
we obtain the following relation 
between $\eta$ and $\theta$: 
\begin{equation}
4\theta\Bigl(\acollata\Bigr)^{-n}\eta^3 -2(1+\theta)\eta +1 = 0 \, .
\label{cubiceq}
\end{equation}

We would like to point out that there is some confusion in the literature
around the exact expression of the cubic equation~(\ref{cubiceq}). 
Some authors (e.g.,  \citealt{iliev}, \citealt{battye}) have derived the same
expression as ours, also using the Poisson equation with the pressure term 
for the dark energy. Other authors, however, 
(e.g., 
\citealt{lokas01}, 
\citealt{basilakos}, 
\citealt{weinberg2003})
seem to have used the approximation of  
$r_{\rm vir}/r_{\rm ta}$ given by \cite{wang}, which is exact for the cosmological
constant case, but differs significantly from the analytical expression 
when $w\neq-1$. Rewriting  equation~(\ref{cubiceq}) using the same notations as
\cite{wang}, we have: 
\begin{equation}
2\eta_v \eta^3 - 2 (1+{{\eta_t}\over{2}})\eta +1 = 0 
\end{equation}
with

\begin{displaymath}
\left\{ 
\begin{array}{l@{=}l}
\eta_t & -{{(n-2)}\over{\zeta}}{{\Omega_{\rm X,ta}}\over{\Omega_{\rm ta}}}  \\
\eta_v &  -{{(n-2)}\over{\zeta}}{{\Omega_{\rm X,vir}}\over{\Omega_{\rm vir}}}  
{ {(1+z_{\rm coll})^3}\over{(1+z_{\rm ta})^3}}. \\
\end{array}
\right  . 
\end{displaymath}

Note that those equations differ from those of \cite{wang} 
where the term $-(n-2)$ in the expressions above of $\eta_t$ and
$\eta_v$ is replaced by 2. The formulae are identical
for $n=0$ (the cosmological constant case) but they differ when $n\neq0$.
This results in different values of the density of the virialised clusters,
as we shall see. 
In particular, using the approximation of \cite{wang} in the $w=-1/3$ case
yields to erroneous results. Because of the contribution of the pressure term in 
the Poisson equation, a factor $(n-2)$ appears in the cubic equation, which
means that for $n=2$ ($w=-1/3$) one should recover the same expression for 
density contrast at virialisation as for an open universe with the same 
$\Omega_0$. This is not the case if the $(n-2)$ term is omitted. 

\subsubsection{The collapse factor versus $\theta$ and $\acollata$}
Let us solve eq.~(\ref{cubiceq}) to express the collapse factor $\eta$. 
For $\theta=0$ (which corresponds to $\rho_X=0$ or to $n=2$),
$\eta=1/2$. 
This is a well-known result for models without dark energy:
the final radius of a virialised sphere is half the turn-around
radius. 
For $\theta\neq0$ one obtains 
a cubic equation in $\eta$ that can be solved analytically 
to express the collapse factor as a function 
of $\theta$ and $\frac{a_{\rm coll}}{a_{\rm ta}}$ for 
different values of $n$.
Here the discriminant of the cubic equation is negative and 
there are three unequal real solutions. 
The first solution is always greater than 1 for the range
of $\theta$ of interest, while the second solution is negative. 
The third solution tends to 1/2 for $\theta\rightarrow0$ and 
it can be expressed as:  
\begin{equation}
\eta = 2\sqrt{\frac{1+\theta}{6\theta}} \Bigl(\acollata\Bigr)^{n/2} 
\cos\left(\frac{\phi+4\pi}{3}\right)
\end{equation}
with 
\begin{equation}
\phi = \arccos\left(-\frac{(6\theta)^{3/2}}{8\theta(1+\theta)^{3/2}} 
\left(\acollata\right)^{-n/2}\right)\, .
\end{equation}
For $w=-1$ (the cosmological constant), $\eta$ is independent of 
$\acollata$ and it can be calculated readily, as was first done 
by \cite{lahav}, who also gave
a simple approximated solution valid for a first-order expansion
of $r_{\rm vir}/r_{\rm ta}$ around 1/2 (the value for $\lambda=0$).

In quintessence models, $\eta$ depends on $\acollata$, which will be
calculated below. 
Fig.~\ref{rhotheta} shows the ratio of the density of the virialised sphere to the
density at maximum expansion versus $\theta$, the strength of the
dark energy at maximum expansion, for different cosmological models. 
In order to reach an equilibrium in the presence of dark energy, 
the sphere has to achieve a higher density as $\theta$ increases. 
The final density of the virialised object is even higher in quintessence
models than in $\Lambda$CDM models, and it increases with increasing
$w$. The curves corresponding to quintessence models are intermediate 
between those
of the $\Lambda$CDM model 
and the open model. 

The minimum radius of the
virialised sphere is reached for $\theta=0.5$. 
In a universe with a cosmological constant, the minimum radius is 
$\eta_{\rm min}=0.366$. 
In quintessence models, the radius of the virialised
sphere tends toward a minimum $\eta_{\rm min}=0.333$  as the ratio
$\acollata$ increases.  
The maximum ratio of the density of the virialised sphere 
to the density at turn-around is thus 
8 for models without dark energy, 
20 for $\Lambda$ models, 
and 27 in quintessence models (when $w\rightarrow -1/3$).  

Expanding $\eta$ around 1/2, its value for $\theta=0$, one obtains a  
first-order approximation:

\begin{equation}
\eta=\frac{1-\theta \left(\frac{a_{\rm coll}}
                          {a_{\rm ta}}\right)^{-n}   }
{2 - \theta \left(
             3  \left(\frac{a_{\rm coll}}{a_{\rm ta}}\right)^{-n} -2
	     \right)} \, .
\end{equation}

\subsubsection{The $\acollata$ ratio}

The $\acollata$ ratio can be derived from the integration of the 
Friedmann equation~(\ref{eq-friedmann}) to calculate the scale factor as a function
of time, using the fact that the collapse
time is twice the turn-around time: 
\begin{equation}
\frac{t_{\rm coll}}{t_{\rm ta}} = 2 \, .
\end{equation}
$\acoll$ and $\ata$ are related by: 
\begin{equation}
\int_0^{a_{\rm coll}} f(a) da = 2 \int_0^{a_{\rm ta}} f(a) da \, .
\end{equation}

Although $\acollata$ is constant for an Einstein-de Sitter 
universe (equal to  $(\frac{t_{\rm coll}}{t_{\rm ta}})^{2/3}=2^{2/3}=1.587$), 
it varies in other cosmologies, decreasing with increasing $z_{\rm coll}$, 
and converges toward the Einstein-de Sitter value at high redshifts.  

\begin{figure}
\psfig{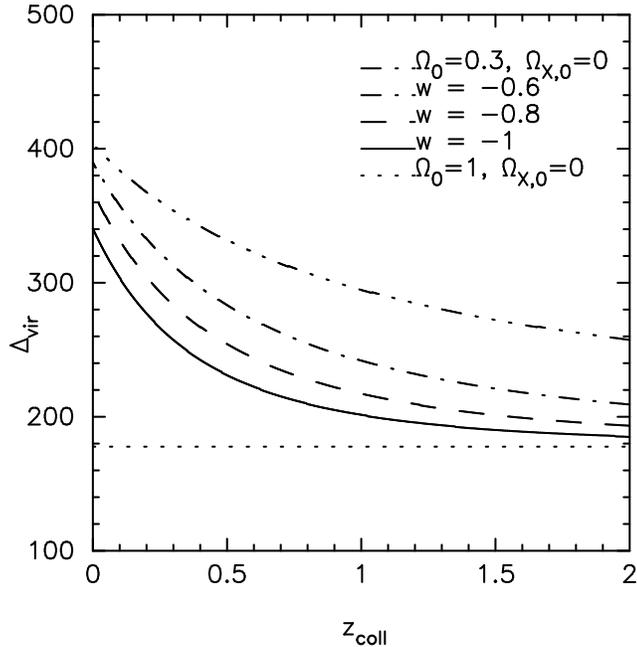}
\caption{
Overdensity $\Delta_{\rm vir}$ 
versus collapse redshift for different cosmological
models. 
Quintessence models are intermediate between 
the $\Lambda$CDM model and the open model. 
}
\label{figdeltavir}
\end{figure}
\subsection{Density contrast at virialisation}  
In Sect.~\ref{sect-expansion} we have calculated the overdensity of the forming
cluster
at turn-around: $\zeta =\left(\rho_{\rm cluster}/\rho_{\rm
bg}\right)_{\rm ta}$. 
In Sect.~\ref{sect-collapse}, we have used the virial theorem to express
$\rho_{\rm vir}/\rho_{\rm cluster,ta}$ as a function of 
$(\rho_X/\rho_{\rm cluster})_{\rm ta}$. 
We are now able to calculate the density contrast of the virialised
cluster 
as a function of the collapse redshift: 
\begin{equation}
\Delta_{\rm vir}(z_{\rm coll}) = \frac{\rho_{\rm vir}}{\rho_{\rm bg,vir}} 
= \eta^{-3} \zeta \left( \frac{1 + z_{\rm ta}}{1 + z_{\rm coll}} \right)^{3} 
\end{equation}
(see 
\citealt{lacey} for a $\Lambda=0$ universe, 
\citealt{eke} for a flat universe with a cosmological constant, 
\citealt{lokaspreprint} for a non-flat universe with $\Lambda$, 
\citealt{wang}  for a flat universe with quintessence). 
The result is shown in Fig.~\ref{figdeltavir}. 
In quintessence models, the overdensity at virialisation takes intermediate values
between the $\Lambda$CDM curve and the open model curve (which is the limiting
case when $w\rightarrow -1/3$). 
The \cite{wang} approximation yields an overestimate of $\Delta_{\rm vir}$
for $w > -1$ by up to 40\% when $w\rightarrow -1/3$. 
\cite{weinberg2003}, for instance, 
find $\Delta_{\rm vir}= 560$ at $z_{\rm coll}=0$ for the
$w=-1/3$ model instead of 400. 
The general form for the expression of the potential energy in the virial 
theorem yields a weaker dependence of $\Delta_{\rm vir}$ on $w$, making
it more difficult to constrain $w$ using weak lensing 
(see also \citealt{bartelmann}). 

\section{Abundance of rich clusters} 
\begin{figure}
\psfig{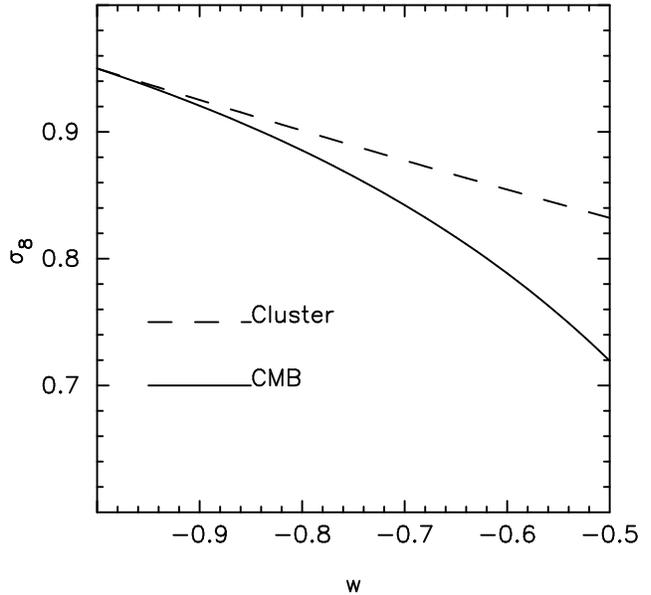}
\caption{Decrease of $\sigma_8$ with increasing $w$, 
using the X-ray cluster normalisation or the CMB normalisation. 
}
\label{figsig8w}
\end{figure}

Let us now calculate the comoving number density 
of virialised clusters 
in a certain mass range 
as a function of redshift: 
\begin{equation}
n(M,z,w) dM = {{\rho_0}\over{M}} 
{{d\nu(M,z,w)}\over{dM}}
f(\nu) dM 
\end{equation}
where $\rho_0$ is the background density at redshift zero, and 
\begin{equation}
  \nu(M,z,w) = {{\delta_c(z,w)}\over{\sigma(M,z,w)}}
\end{equation}
where $\delta_c$ is the linear overdensity of a perturbation collapsing at 
redshift $z$ and 
$\sigma$ is the r.m.s. of the mass fluctuation in spheres containing the mass $M$. 
$\delta_c$ depends weakly on the cosmological model and is close to the 
Einstein-de Sitter value of 1.68. We have taken the fit provided by \cite{weinberg2003}
to $\delta_c$ as a function of $z$ and $w$. 

The standard Press-Schechter mass function is of the form: 
$f(\nu) = \sqrt{{{2}\over{\pi}}} \exp{(-\nu^2/2)}$. 
It provides a good general representation of the observed distribution of clusters.
However, it is known to predict too many low-mass clusters and too few high mass
clusters, as well as too few clusters at high redshifts. 
We use the modified Press-Schechter function 
proposed by  Sheth \& Tormen (1999) 
that gives 
substantially better fits to simulated mass functions 
(\citealt{jenkins}). 
This formula has been given a physical justification in terms of an ellipsoidal model for 
the collapse of perturbations  
(\citealt{smt01}). The mass function is
\begin{equation}
f(\nu) = \sqrt{{{2}\over{\pi}}} 0.2709 (1 + 1.1096 \nu^{0.6})\exp{(-0.707\nu^2/2)}
\end{equation}

We need to calculate the amplitude of the mass fluctuations $\sigma(M)$, as well 
as its derivative with respect to $M$. 
In a Gaussian density field, the variance depends on the scale $R$: 

\begin{equation}
\sigma^2(R)= {{1}\over{2\pi^2}} \int_0^\infty 
k^3 P(k) W^2(kR) {{dk}\over{k}} \, , 
\end{equation}
where $R(M)= \left( {{3M}\over{4\pi\rho_0}} \right)^{1/3}$
is the Lagrangian radius of a halo of mass $M$ at the present time, 
$W(u) = 3\left(\sin(u) -u \cos(u)\right)/u^3$ is the Fourier transform of 
a spherical top-hat filter with radius $R$, and $P(k)$ is the power spectrum
of density fluctuations extrapolated to $z=0$ according to linear theory. 
Assuming that the baryon density parameter $\Omega_{B,0} \ll \Omega_{\rm CDM,0}$, 
the CDM power spectrum can be approximated by
$P(k) \propto k T^2(k)$.
As for the transfer function,  we have used

\begin{displaymath}
\begin{array}{ccl}
T(k) & = & {{\ln(1+2.34q)}\over{2.34 q }} \\
     &\times
& [1 + 3.89 q + (16.1 q )^2 + (5.46 q)^3 + (6.71 q)^4
]^{-1/4}
\end{array}
\end{displaymath}
with $q= k/[\Omega_0 h^2 {\rm Mpc}^{-1}]$ \citep{bardeen}. 

We normalise the power spectrum by setting the value of the amplitude 
of the fluctuation within $8 h^{-1} {\rm Mpc}$ spheres in a
$\Lambda$CDM model. 

$\sigma_8$ can be constrained from observations of the cosmic microwave
background (CMB). 
\cite{doran} provide a useful approximation relating 
$\sigma_8$ in a universe with quintessence to the
$\sigma_8$ in a corresponding $\Lambda$CDM universe with the same 
quantity of dark energy, and consistent with the COBE normalisation
(\citealt{bunn}). 
In the case of a dark energy component with 
a constant equation of state, as in the models we consider here, 
the expression is  simple and 
precise to about 5\% : 
\begin{equation}
{{\sigma_8^X}\over{\sigma_8^\Lambda}} \simeq
(1-\Omega_{X,0})^{-(1+w)/5} \sqrt{ 
{{\tau_0^X}\over{\tau_0^\Lambda}}
}
\end{equation}
where $\tau_0=\tau(a=1)$ is the conformal age of the universe: 
$\tau(a) = \int_0^a {{da'}\over{a'^2 H(a')}}$. 

$\sigma_8$ can also be inferred from the observed abundance 
of X-ray clusters (\citealt{wang}):
\begin{equation}
\sigma_8 \simeq 0.5 \Omega_0^{-\gamma(\Omega_0,w)} 
\end{equation}
where $\gamma$ is a function of $\Omega_0$ and $w$. 

In Fig.~\ref{figsig8w}, 
we show both normalisations of $\sigma_8$: the CMB and the
X-ray cluster normalisation. 
We have taken $\sigma_8^\Lambda= 0.95$. 
$\sigma_8$ has to decrease with increasing $w$ 
to meet the observational constraints. 
The CMB normalisation implies a steeper decrease of $\sigma_8$
with increasing $w$. 

The cluster evolution is sensitive both to $\sigma_8$ and 
to $w$ and might provide
a way to constrain those important parameters. However, the 
degeneracy between $w$ and $\sigma_8$ produces a near cancellation
of the two effects. We have shown that, as $w$ increases, 
structures form earlier and are more concentrated. However, 
because of the accompanying decrease in $\sigma_8$, more clusters
will form  at a later epoch. 

Now we want to calculate a more directly observable quantity, namely 
the number density of clusters above a certain mass 
per unit steradian and 
per redshift interval:  
\begin{equation}
{{d^2N(z)}\over{d\Omega dz}} = 
{{d^2V_c}\over{d\Omega dz}}
\int_{M_{\rm lim}}^{+\infty} n(M,z) dM
\end{equation}
where the comoving volume is 
\begin{equation}
{{d^2V_c}\over{d\Omega dz}} = c{{d_A(z)^2 (1 + z)^2}\over{H(z)}}
\end{equation}
and $d_A$ is the angular diameter distance. 

\begin{figure}
\psfig{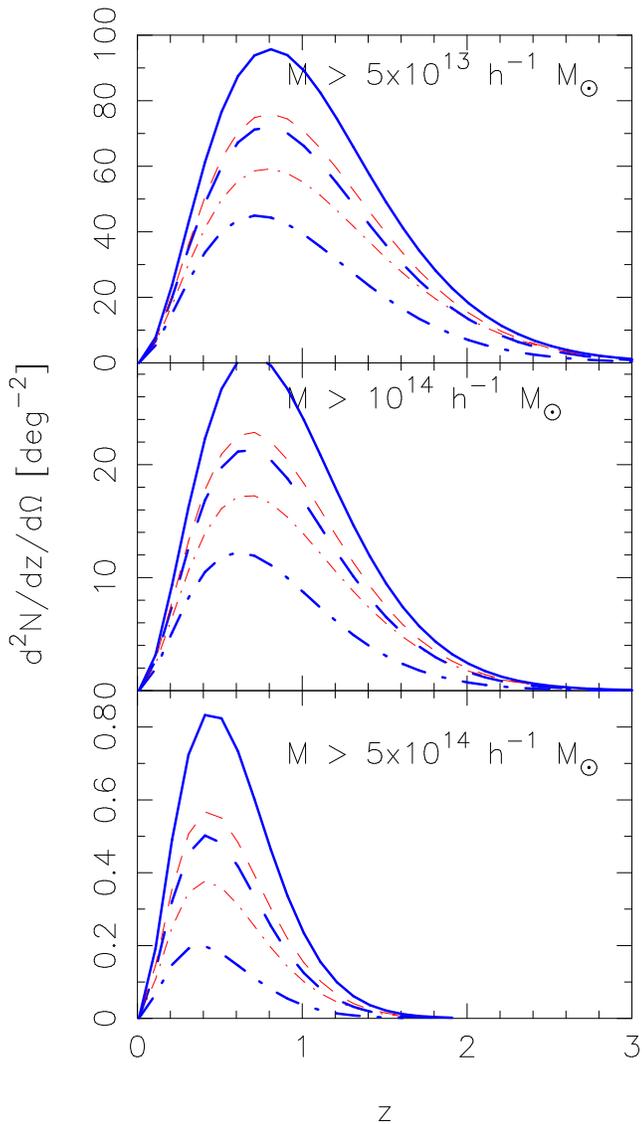}
\caption{Number density of virialised clusters 
with mass larger than a given mass threshold
per redshift interval and square degree in the three cosmological  
models with dark energy, using the X-ray cluster normalisation (thin lines)
or the CMB normalisation (thick lines).  
The solid line shows the $\Lambda$CDM, the dashed line the $w=-0.8$ model   
and the dotted-dashed line the $w=-0.6$ model. 
}
\label{figd2N}
\end{figure}

Fig.~{\ref{figd2N}} shows that quantity as a function of 
redshift for the different dark energy models and
using both the CMB and the cluster $\sigma_8-w$ relations. 
As the threshold mass of the cluster catalog increases, 
the peak of the redshift distribution is shifted toward
lower redshifts and the number density of clusters 
decreases. The amplitude of the curves is sensitive
to the $\sigma_8-w$ relation, but not the position of 
the peak in redshift. 
Note that we have assumed a constant limiting mass as a function
of redshift. 
It is possible to carry out a similar calculation assuming a dependence 
of $M_{\rm lim}$ with 
$z$, appropriate for a given survey. 
For instance the expected yield of a survey of galaxy clusters using the
Sunyaev-Zeldovich effect is sensitive to the mass threshold of detection 
(e.g., \citealt{holder}). 
Also, such an approach of course assumes 
that the cluster masses are known. 
In practice, the total masses have to be inferred assuming some scaling
relation with an observable (for instance the temperature of the X-ray
emitting gas). The evolution of the mass-observable relation has to be
well-calibrated. In principle, this can be achieved in 
dark energy models with constant $w$ if the survey is deep enough so 
that the clusters counts can be binned both in redshift and in the observable
quantity \citep{hu03}. 

\section{Summary and conclusion}

Using the top-hat model 
and the Sheth-Tormen refinement of the Press-Schechter mass 
function,  
we have examined the effect of the equation of state of dark energy 
on the evolution of spherical overdensities and the abundance of 
clusters. 
We have considered constant-$w$ models. 
We have used a general expression for the virial theorem in the presence of 
dark energy instead of the \cite{wang} approximation widely used in the literature and 
valid mainly around $w=-1$. 
Using the general expression, one can show that a dark energy model with an equation
of state $w=-1/3$ is equivalent to an open CDM model with the same matter density
$\Omega_0$ and no dark energy. 
The \cite{wang} approximation in that regime yields  
an overestimate of the overdensity of the virialised cluster $\Delta_{\rm vir}$ by
as much as 40\%.   
A flat dark energy model with $w=-1/3$ differs, however, from the corresponding
open CDM model because
of the different curvature, implying a different distance-redshift relation and 
different number densities of virialised clusters per redshift interval and per
square degree on the sky. 

The number density of massive clusters is sensitive 
both to the amplitude of the mass fluctuations, 
$\sigma_8$, 
and to the $\sigma_8-w$ relation. 
As $w$ increases, $\sigma_8$ has to decrease to match the 
observations of the cosmic microwave background and the observed abundance
of X-ray clusters. 
As $w$ increases, clusters form earlier and are more concentrated. 
However, this effect
is partly counteracted by the decrease of $\sigma_8$ with $w$. 
A recent analysis of the evolution of the X-ray temperature of clusters 
favours a low $\sigma_8$ and a higher $w$: 
$\sigma_8= 0.66\pm0.16$  with $w=-(0.42\pm0.21)$ at 68\% confidence level
(\citealt{henry04}). 
A previous similar analysis with fewer distant clusters in a flat
$\Lambda$CDM model had yielded a higher value of $\sigma_8$ 
($0.77\pm0.15$, \citealt{henry00}). 
Breaking the $\sigma_8-w$ degeneracy will require 
very sensitive observations of the evolution of the cluster abundance  
together with a good knowledge 
of the mass-temperature relation and 
the limiting mass of the survey. 

\section*{Acknowledgments}
This paper has benefited from a discussion with Alexandre Refregier 
about the growth factor in quintessence models. 
We are grateful to John Black for useful comments on the manuscript. 
C.H. is grateful to John Bahcall for an invitation to visit the Institute
for Advanced Study in Princeton in September 2004 where part of this work 
was completed. 
C.H. acknowledges financial support from the Swedish Research Council 
{\it Vetenskapsr\aa det}.

\label{lastpage}


\begin{thebibliography}{99}

\bibitem[\protect\citeauthoryear{Bardeen et al.}{1986}]{bardeen} 
Bardeen J.M., Bond J.R., Kaiser N., Szalay A.S., 1986, ApJ 304, 15
\bibitem[\protect\citeauthoryear{Bartelmann, Perrotta \& Baccigalupi}{2002}]{bartelmann} Bartelmann M., Perrotta F., Baccigalupi C., 2002, A\&A 396, 21, see also 2003, A\&A 400, 19 (erratum)
\bibitem[\protect\citeauthoryear{Basilakos}{2003}]{basilakos} Basilakos S., 2003, ApJ 590, 636 
\bibitem[\protect\citeauthoryear{Battye \& Weller} {2003}] {battye} Battye R.A., Weller J., 2003, Phys. Rev. D 68, 083506
\bibitem[\protect\citeauthoryear{Bildhauer, Buchert \& Kasai}{1992}]{bildhauer} Bildhauer S., Buchert T., Kasai M., 1992, A\&A 263, 23
\bibitem[\protect\citeauthoryear{Bunn \& White}{1997}]{bunn} Bunn E.F., White M.J.,
1997, ApJ 480, 6
\bibitem[\protect\citeauthoryear{Caldwell, Dave \& Steinhardt}{1998}]{caldwell} Caldwell R.R., Dave R., Steinhardt P.J., 1998, Phys. Rev. Lett. 80, 1582
\bibitem[\protect\citeauthoryear{Carroll, Press \& Turner}{1992}]{carroll} Carroll S.M., Press W.H., Turner E.L., 1992, ARAA 30, 499
\bibitem[\protect\citeauthoryear{Chernin, Nagirner \& Starikova}{2003}]{chernin} Chernin A.D., Nagirner D.I., Starikova S.V., 2003, A\&A 399, 19
\bibitem[\protect\citeauthoryear{Doran, Schwindt \& Wetterich}{2001}]{doran} 
Doran, M., Schwindt, J.-M., Wetterich, C., 2001, Phys. Rev. D 64,
123520 
\bibitem[\protect\citeauthoryear{Eke, Cole \& Frenk}{1996}]{eke} Eke V.R., Cole S., Frenk C.S., 1996, MNRAS 282, 263
\bibitem[\protect\citeauthoryear{Gunn \& Gott}{1972}]{gunn} Gunn J.E., Gott J.R., 1972, ApJ 176, 1
\bibitem[\protect\citeauthoryear{Hamilton}{2001}]{hamilton} Hamilton A.J.S., 2001, MNRAS 322, 419
\bibitem[\protect\citeauthoryear{Henry}{2000}]{henry00} Henry, J.P., 2000, ApJ 534, 565 
\bibitem[\protect\citeauthoryear{Henry}{2004}]{henry04} Henry, J.P., 2004, ApJ 609, 603
\bibitem[\protect\citeauthoryear{Holder et al.}{2000}]{holder} Holder G.P., Mohr J.J., Carlstrom J.E., 
Evrard A.E., Leitch E.M., 2000, ApJ 544, 629 
\bibitem[\protect\citeauthoryear{Hu}{2003}]{hu03} Hu W., 2003, Phys. Rev. D 67, 081304
\bibitem[\protect\citeauthoryear{Iliev \& Shapiro}{2001}]{iliev} Iliev I.T., Shapiro P.R., 2001, MNRAS 325, 468
\bibitem[\protect\citeauthoryear{Jenkins et al.}{2001}]{jenkins} Jenkins A., Frenk C.S., White S.D.M., 
Colberg J.M., Cole S., Evrard A.E., Couchman H.M.P., Yoshida N.,  
2001, MNRAS 321, 372
\bibitem[\protect\citeauthoryear{Lacey \& Cole}{1993}]{lacey} Lacey C., Cole S., 1993, MNRAS 262, 627
\bibitem[\protect\citeauthoryear{Lahav et al.}{1991}]{lahav} Lahav O., Lilje P.B., Primack J.R., Rees M.J., 1991, MNRAS 251, 128
\bibitem[\protect\citeauthoryear{Landau \& Lifshitz}{1960}]{landau} Landau L.D., Lifshitz E.M., 1960, Mechanics, Pergamon Press, Oxford
\bibitem[\protect\citeauthoryear{Lilje}{1992}]{lilje92} Lilje P., 1992, ApJ 386, L33
\bibitem[\protect\citeauthoryear{Linder \& Jenkins}{2003}]{linderjenkins} Linder E.V., Jenkins A., 
2003, MNRAS 346, 573
\bibitem[\protect\citeauthoryear{{\L}okas \& Hoffman}{2001}]{lokaspreprint} {\L}okas E.L., Hoffman Y., 2001, preprint, astro-ph/0108283 
\bibitem[\protect\citeauthoryear{{\L}okas}{2001}]{lokas01} {\L}okas E.L.,
2001, Acta Physica Polonica B 32, 3643
\bibitem[\protect\citeauthoryear{{\L}okas, Bode \& Hoffman}{2004}]{lokas04} {\L}okas E.L., Bode P., Hoffman Y.
2004, MNRAS 349, 595 
\bibitem[\protect\citeauthoryear{Mota \& van de Bruck}{2004}]{mota} Mota D.F., van de Bruck C., 2004, A\& A 421, 71 
\bibitem[\protect\citeauthoryear{Peebles}{1980}]{peebles} Peebles P.J.E., 1980, The Large-Scale Structure of the Universe. Princeton University Press, Princeton NJ 
\bibitem[\protect\citeauthoryear{Peebles \& Ratra}{1988}]{peeblesratra88} Peebles P.J.E., Ratra B., 1988 ApJ Letters 325, L17 
\bibitem[\protect\citeauthoryear{Peebles \& Ratra}{2003}]{peeblesratra03} Peebles P.J.E., Ratra B., 
2003, Rev. Mod. Phys. 75, 559
\bibitem[\protect\citeauthoryear{Press \& Schechter}{1974}]{press74} Press W.,
Schechter P., 1974, ApJ 187, 425 
\bibitem[\protect\citeauthoryear{Ratra \& Peebles}{1988}]{ratrapeebles88} Ratra B., Peebles P.J.E., 1988, Phys. Rev. D 37, 3406  
\bibitem[\protect\citeauthoryear{Sheth \& Tormen}{1999}]{st99} Sheth R.K., Tormen G., 
1999, MNRAS 308, 119
\bibitem[\protect\citeauthoryear{Sheth, Mo \& Tormen}{2001}]{smt01} 
Sheth R.K., Mo H.J., Tormen G., 2001, MNRAS, 323, 1 
\bibitem[\protect\citeauthoryear{Viana \& Liddle}{1996}]{vianaliddle} Viana P.T.P., Liddle A.R., 1996, MNRAS 281, 323
\bibitem[\protect\citeauthoryear{Wang \& Steinhardt}{1998}]{wang} Wang L., Steinhardt P.J., 1998, ApJ 508, 483
\bibitem[\protect\citeauthoryear{Weinberg \& Kamionkowski}{2003}]{weinberg2003} Weinberg N.N., Kamionkowski M., 2003, MNRAS 341, 251
\end{thebibliography}
\end{document}